\documentclass{PoS}

\usepackage{listings}

\newif\ifpreprint
\preprinttrue

\title{
\ifpreprint
\hbox{\rm\small
FR-PHENO-2014-006$\null\hskip 3.7cm \null$
IPhT--t14/092$\hskip 4.5cm \null$ 
IPPP/14/62$\null\hskip 2.4cm \null$
\break}
\hbox{\rm\small 
LPN14-081 $\null\hskip 1.9cm \null$
SB/F/440--14$\null\hskip 1.9cm \null$
SLAC--PUB--16006$\null\hskip 1.8cm \null$
UCLA-14-TEP-105$\null\hskip 1.2cm \null$
\break}
\hbox{$\null$\break}
\fi
High multiplicity processes with BlackHat and Sherpa}

\ShortTitle{High multiplicity processes with BlackHat and Sherpa}

\author{Zvi Bern, Kemal Ozeren\\
        Department of Physics and Astronomy, UCLA, 
		Los Angeles, CA 90095-1547, USA\\
        E-mail: \email{bern@physics.ucla.edu}, \email{ozeren@physics.ucla.edu}
		}
\author{Stefan H\"oche\\
        SLAC National Accelerator Laboratory, Stanford University,
        Stanford, CA 94309, USA\\
        E-mail: \email{lance@slac.stanford.edu}, \email{shoeche@slac.stanford.edu}
		}
\author{Fernando Febres Cordero\\
  		Departamento de F\'{\i}sica, Universidad
		Sim\'on Bol\'{\i}var, Caracas 1080A, Venezuela\\
        E-mail: \email{ ffebres@usb.ve}
		}
\author{Harald Ita\\
		Albert-Ludwigs-Universit\"at Freiburg, 
		Physikalisches Institut,
		D-79104 Freiburg, Germany\\
        E-mail: \email{harald.ita@physik.uni-freiburg.de}
		}
\author{David Kosower, Nicola Adriano Lo Presti\\
        Institut de Physique Th\'eorique, CEA--Saclay,
        F--91191 Gif-sur-Yvette cedex, France\\
        E-mail: \email{david.kosower@cea.fr}, \email{nicola.lo-presti@cea.fr} 
	}
\author{Stephanie Bartle, Jeppe R. Andersen, \speaker{Daniel Ma\^{\i}tre}
	\\
        Institute for Particle Physics Phenomenology,
        Department of Physics, University of Durham,
        DH1 3LE, UK\\
        E-mail: \email{stephanie.bartle@durham.ac.uk}, \email{jeppe.andersen@durham.ac.uk}, \email{daniel.maitre@durham.ac.uk}
}

\abstract{ In this contribution, we present an intermediate storage
   format for next-to-leading order
  (NLO) events and explain the advantages of presenting a NLO calculation in
  this format. We also present some recent applications, including the
  calculation of PDF uncertainties and the combination of different
  multiplicity samples for the prediction of gap fractions in inclusive
  dijet events.  }

\FullConference{ Loops and Legs in Quantum Field Theory - LL 2014,\\
		 27 April - 2 May 2014 \\
		 Weimar, Germany }

\def\ntuple{$n$-tuple}

\begin{document}

\section{Introduction}
In recent years much progress has been achieved in the calculation of QCD
predictions to next-to-leading order (NLO) accuracy (for a summary, see
ref.~\cite{Butterworth:2014efa}).  Even with these advances, high-multiplicity
processes, though now feasible, remain computationally expensive. In this
contribution, we report on a method of using specialized event files,
which we call \ntuple{} files,
to reduce the cost of fixed-order NLO
calculations.  This is important as more and more techniques such as NLO
parton-shower matching and merging use them as an input. 
For example, the event files
described in this contribution
 have been used within the LoopSim method
\cite{Rubin:2010xp} to merge NLO calculations for Z+1 jet and Z+2 jets
\cite{Maitre:2013wha}. In the next section we describe the \ntuple{} files
and a library for their use. In the third section we show some applications of
the \ntuple{} files.

\section{n-Tuples}
Next-to-Leading Order (NLO) calculations are computationally intensive, which
means that under normal circumstances computing new observables with new cuts
is a tedious task. The most computationally demanding part is the calculation
of the matrix elements; other operations such as jet clustering, the
evaluation of the parton distribution functions and the calculation of the
observables are relatively cheap. We can amortize the cost of the matrix
element calculation by storing the matrix elements and the phase-space
information along with a few coefficients of the logarithms driving the scale
dependence in a file. These files can then be re-read to obtain an analysis
with different cuts or observables, or to yield the result one would have
obtained with a different PDF or a different choice of renormalization or
factorization scale.  This is especially useful when computing PDF
uncertainties that would otherwise require the same matrix element to be
recomputed a large number of times.

The ROOT \cite{ROOT} format has been chosen as a backend to store the matrix
elements and associated information. Table \ref{tab:ntuples} details the
hadronic center of mass energy and minimum transverse momentum cuts applied on the jets for each processes available in the \ntuple{} format. Details about the calculation for the creation of these
files can be found in refs.~\cite{W3PRL,W3,W4,ItaOzeren,Z3,Z4,pureQCD}.

NLO event files such as the one we describe here and in ref. \cite{Bern:2013zja} have the added advantage of making it easier to communicate challenging NLO computations with the experimental community.

Along with the \ntuple{} files we provide a C++ library for accessing the
information they contain. It can either be used out of the box or as a
template for a dedicated implementation in a different framework. The
library also provides a python interface. Figure \ref{fig:ex1} shows an
example of the usage of the library to read a \ntuple{} file (and in this case
display the stored momenta instead of using them to compute an observable or
verifying that this particular event passes the analysis cuts.) Figure
\ref{fig:ex2} shows an example of the usage of the library to change the
factorization and renormalization scale for a new prediction.

In t next section we present some applications of this method that would have
been too prohibitive in CPU time to perform using straightforward 
repeated evaluation of the matrix elements.

\begin{figure}
\begin{lstlisting}[basicstyle=\ttfamily,keepspaces=true,columns=fullflexible,escapeinside={(*}{*)}]
import nTupleReader as NR
r=NR.nTupleReader()

r.addFile("sample.root")

while r.nextEntry():
	for i in range(r.getParticleNumber()):
		print "p(%d)=(%f,%f,%f,%f)" % (
	 	 i,
	 	 r.getEnergy(i),
	 	 r.getX(i),
	 	 r.getY(i),
	 	 r.getZ(i)
		)
\end{lstlisting}
\caption{Example of the usage of the nTupleReader library. The example uses the python interface of the library.}\label{fig:ex1}

\end{figure}
\begin{figure}
\begin{lstlisting}[basicstyle=\ttfamily,keepspaces=true,columns=fullflexible,escapeinside={(*}{*)}]
import nTupleReader as NR
r=NR.nTupleReader()
r.addFile("sample.root")

r.setPDF("CT10nlo.LHgrid")
r.setPDFmember(12)

while r.nextEntry():
	# compute new scales
	RenScale = ....
	FacScale = ....
	newWeight=r.computeWeight(FacScale,RenScale)
	# use this weight in the analysis
	...
\end{lstlisting}
\caption{Example of the usage of change of scales using the nTupleReader library. The example uses the python interface of the library.}\label{fig:ex2}
\end{figure}

\begin{table}
\begin{tabular}{|c|c|l|}
\hline
Process & energy & pt cut \\
\hline
$W^+(\rightarrow e^+\nu_e)+1,2,3,4\mbox{ jets }$ & 7~TeV & 25~GeV
\\
$W^+(\rightarrow e^+\nu_e)+1,2,3\mbox{ jets }$ & 8~TeV & 20~GeV
\\ 
$W^{-}(\rightarrow e^-\bar \nu_e)+1,2,3,4\mbox{ jets }$ & 7~TeV & 25~GeV 
\\
$W^{-}(\rightarrow e^-\bar \nu_e)+1,2,3\mbox{ jets }$ & 8~TeV & 20~GeV
\\
$Z(\rightarrow e^+e^-)+1,2\mbox{ jets }$ & 7~TeV & 25~GeV
\\
$Z(\rightarrow e^+e^-)+3,4\mbox{ jets }$ & 7~TeV & 20~GeV
\\
$Z(\rightarrow e^+e^-)+1,2,3\mbox{ jets }$ & 8~TeV & 20~GeV
\\
$n\mbox{ jets }$ ($n=1,2,3,4$) & 7~TeV & 40~GeV
\\
$n\mbox{ jets }$ ($n=1,2,3,4$) & 8~TeV & 40~GeV
\\
\hline
\end{tabular} 
\caption{Available processes at NLO. The decay of the vector boson into a lepton pair is always included.}
\label{tab:ntuples}
\end{table}

\section{Applications}
\subsection{Jet Veto}
Understanding the impact of jet vetoes is very important for current Higgs
studies. The behavior of observables when a jet veto is applied can be
investigated in processes that are under better theoretical control than
Higgs production. New calculations or techniques can be checked against data for simpler processes and that knowledge can be used to improve the our
understanding of Higgs or BSM measurements. One interesting observable for the investigation
of jet veto efficiencies is the gap fraction $g$, which is defined as the
probability of having no jet above a threshold $Q_0$ between the two tagging
jets:
\[g=\frac{\sigma_{Y/pt}(Q_0)}{\sigma_{tot}}\] 
where $\sigma_{Y/p_T}(Q_0)$ is the cross section when vetoing jets with both
transverse momentum above the threshold $Q_0$ and rapidity between that of
the tagging jets. These can be either the two highest transverse momentum
jets ($\sigma_{p_T}$) or the most forward/backward ones
($\sigma_Y$). $\sigma_{tot}$ is the total cross section without the jet
veto. We use the notation
\[\sigma_{g=n}\;,\;\sigma_{g\geq n}\]
to denote the cross section with exactly $n$ jets in the gap , or $n$ jets or
more in the gap, respectively.

\begin{figure}
\includegraphics[scale=0.50]{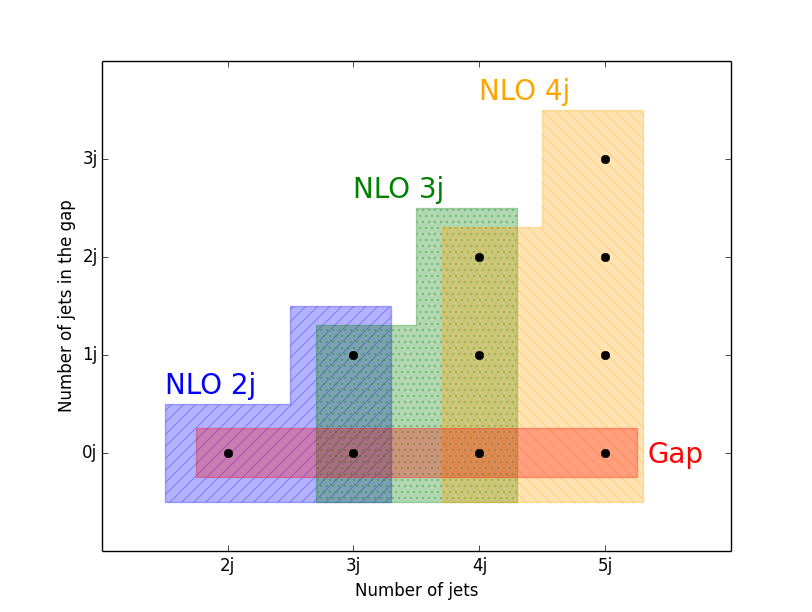}
\caption{Each dot represent a possible contribution.}
\label{fig:gap}
\end{figure}

Restricting the precision of the fixed order prediction to NLO, one can give a prediction for the gap fraction in two different ways. First, one could use only one NLO calculation for each of the numerator or denominator:
\begin{eqnarray}\label{gapNLO}
g&=&\frac{\sigma_{g=0}}{\sigma_{tot}}
=1-\frac{\sigma_{g\ge 1}}{\sigma_{tot}}\nonumber\\
&=&1-\frac{\sigma^{{\rm nlo},j\geq 3}_{g\geq 1}}{\sigma^{{\rm nlo},j\geq 2}}
=1-\frac{\sigma^{{\rm nlo},j\ge 3}-\sigma^{{\rm nlo},j=3}_{g=0}-\sigma^{{\rm lo},j=4}_{g=0} }{\sigma^{{\rm nlo},j\geq 2}}\;.
\end{eqnarray} 
Alternatively one can try to use more NLO calculations
\begin{eqnarray}\label{gapExSum}
g=\frac{\sigma_{g=0}}{\sigma_{tot}}
=\frac{\sigma^{{\rm nlo},j=2}_{g=0}+\sigma^{{\rm nlo},j=3}_{g=0}+\sigma^{{\rm nlo},j\geq 4}_{g=0}}{\sigma^{{\rm nlo},j=2}+\sigma^{{\rm nlo},j=3}+\sigma^{{\rm nlo},j\geq 4}}\;.
\end{eqnarray}
The two formulae are formally of the same order. Figure \ref{fig:gap}
illustrates the different contributions in the plane spanned by the number of
jets and the number of jets in the gap. Each term in the formulae above can be
identified in this plane.

\begin{figure}
\includegraphics[scale=0.50]{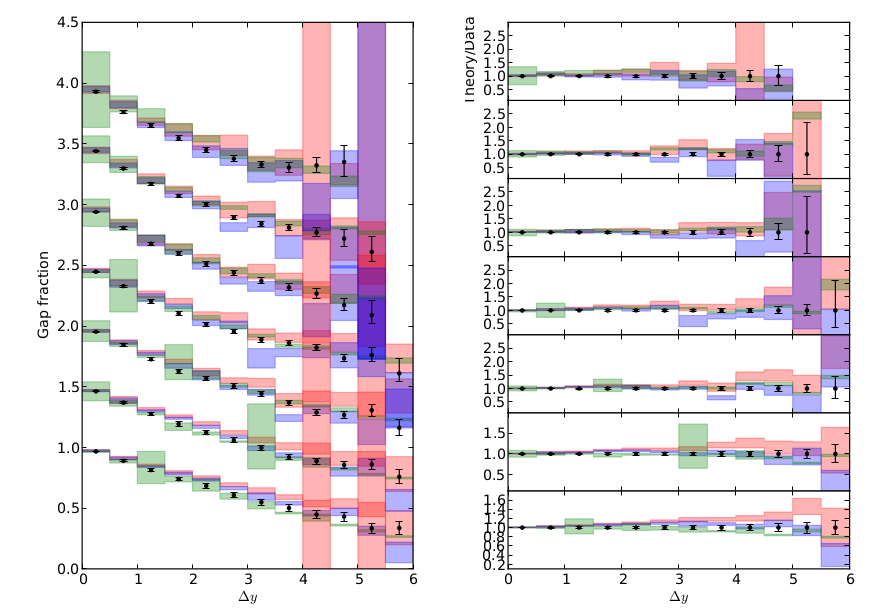}
\caption{
Gap fraction as a function of $\Delta y$ for various slices of $\bar p_T$. The jets defining $\bar p_T$ and $\Delta y$ are the two jets with the largest $p_T$. }
\label{fig:gapfraction}
\end{figure}

The left pane of figure \ref{fig:gapfraction} shows the gap fraction as a
function of the rapidity separation of the tagging jets. The
different sets of curves correspond to different bins in the average
transverse momentum of the two tagging jets $\bar p_T$. Each set of curves is
offset with respect to the lower $\bar p_T$ set by $0.5$. The bins are
\begin{eqnarray}\label{ptSlices}
240\,{\rm GeV}< &\bar p_T& < 270\, {\rm GeV}\nonumber\\
210\,{\rm GeV}< &\bar p_T& < 240\, {\rm GeV}\nonumber\\
180\,{\rm GeV}< &\bar p_T& < 210\, {\rm GeV}\nonumber\\
150\,{\rm GeV}< &\bar p_T& < 180\, {\rm GeV}\nonumber\\
120\,{\rm GeV}< &\bar p_T& < 150\, {\rm GeV}\nonumber\\
90\,{\rm GeV}< &\bar p_T& < 120\, {\rm GeV}\nonumber\\
70\,{\rm GeV}< &\bar p_T& < 90\, {\rm GeV}
\end{eqnarray}
The data points are from the ATLAS measurement \cite{Aad:2011jz}. We provide
theoretical predictions obtained using HEJ
\cite{Andersen:2009nu,Andersen:2009he,Andersen:2011hs} and NLO predictions
obtained by BlackHat+Sherpa
\cite{BH,pureQCD,sherpa1,sherpa2,amegicI,amegicII}.

The right pane of figure \ref{fig:gapfraction} shows the ratio to the data
for each $\bar p_T$ bin. The green band represents the HEJ prediction
while the blue and red curves correspond to the NLO predictions of formulae
(\ref{gapNLO}) and (\ref{gapExSum}), respectively. The bands represent only
the
statistical Monte Carlo integration errors.

\subsection{PDF uncertainties}
\begin{figure}
\includegraphics[scale=0.50]{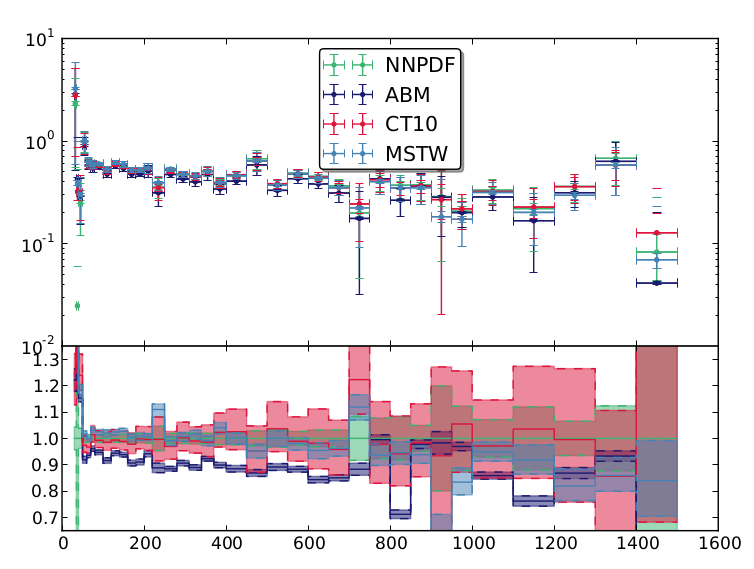}
\caption{$W^{-}$+4 jets to $W^+$ + 4 jets ratio for the first jet transverse momentum for different PDF sets. The lower pane shows the ratio to the NNPDF prediction. The different color bands display the uncertainties.}
\label{fig:WmWp}
\end{figure}

\begin{figure}
\includegraphics[scale=0.50]{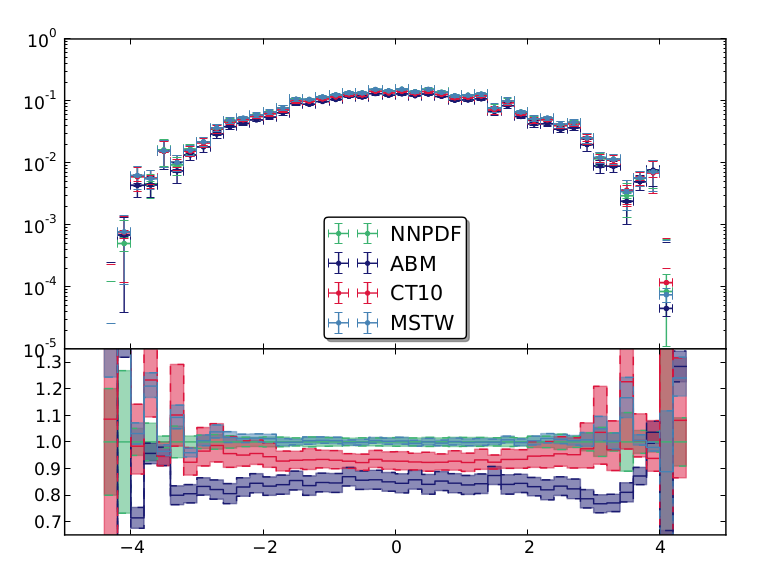}
\caption{Second jet rapidity in Z+4 jets events.}
\label{fig:Z4}
\end{figure}

PDF uncertainties are usually computationally intensive, as the same calculation
has to be preformed with a large number of slightly different PDF fits. Using
\ntuple{} files the expensive part of the calculation need be performed only
once (and in this case it had been done previously, so we get the results at
almost no computational cost). Figure \ref{fig:WmWp} shows the ratio of the
first jet transverse momentum in $W^-$+4 jets and $W^+$+ 4 jets. This ratio
is evaluated for different PDFs and the associated uncertainties are shown
in the lower pane. We have used the NNPDF21 \cite{Ball:2011mu}, MSTW2008
\cite{Martin:2009iq}, CT10 \cite{Lai:2010vv} and ABM11 \cite{Alekhin:2012ig}
PDF sets. The bands for NNPDF provides the 1-$\sigma$ error bands, for MSTW2008 we used the 68\% confidence
level uncertainty estimate, for CT10 the bands represent the 90\% confidence level uncertainty estimate. The errors provided with the ABM set represent a 1-$\sigma$ deviation from the best fit. Figure \ref{fig:Z4} shows the rapidity of the second jet in
events with a Z boson and 4 jets.


\section{Conclusions}
In this contribution we have described a format for NLO events and shown some applications.
\section*{Acknowledgments}
This research was
supported by the US Department of Energy under contracts
DE--AC02--76SF00515 and DE-FG02-13ER42022.  DAK and NALP's research is
supported by the European Research Council under Advanced Investigator
Grant ERC--AdG--228301.  DM's work was supported by the Research
Executive Agency (REA) of the European Union under the Grant Agreement
number PITN--GA--2010--264564 (LHCPhenoNet). SH's work was partly
supported by a grant from the US LHC Theory Initiative through NSF
contract PHY--0705682.  This research used resources of Academic
Technology Services at UCLA, and of the National Energy Research
Scientific Computing Center, which is supported by the Office of
Science of the U.S. Department of Energy under Contract
No.~DE--AC02--05CH11231.

\bibliographystyle{JHEP}
\bibliography{ll2014}

%

\end{document}